\documentstyle[twoside,fleqn,espcrc2]{article}

\title {Operator ordering for generally covariant systems}

\author{Rafael Ferraro\address{Departamento de F\'\i sica,
Facultad de Ciencias Exactas y Naturales, Universidad de Buenos Aires,\\ Ciudad
Universitaria, Pabell\' on I, 1428 Buenos Aires, Argentina}$^{,b}$\thanks{Electronic
address: ferraro@iafe.uba.ar}
 and Daniel M. Sforza\address{Instituto de
Astronom\'\i a y F\'\i sica del Espacio, Casilla de Correo 67 - Sucursal 28, 1428 Buenos
Aires, Argentina}\thanks{Electronic address: sforza@iafe.uba.ar}}

\begin{document}

\begin{abstract}
The constraint operators belonging to a generally covariant system are found out within
the framework of the BRST formalism. The result embraces quadratic Hamiltonian constraints
whose potential can be factorized as a never null function times a gauge invariant
function. The building of the inner product between physical states is analyzed for
systems featuring either intrinsic or extrinsic time.
\end{abstract}

\maketitle

An essential aspect of a generally covariant system is the invariance of its action under
reparametrizations; this means that the label that parametrizes the trajectories of the
system is not the time but a physically irrelevant parameter. As a consequence, the system
is constrained to remain on the hypersurface of the phase space where the Hamiltonian is
null. In fact, since the ``evolution" generated by the Hamiltonian can be regarded as a
reparametrization of the classical trajectory, then the Hamiltonian behaves like a
generator of a gauge transformation of the system; so the Hamiltonian is a first class
constraint. Besides the Hamiltonian constraint ${\cal H}_o$ associated with the
reparametrization invariance, the system can exhibit additional gauge invariance generated
by first class constraints ${\cal H}_a$ linear and homogeneous in the momenta, telling
that some canonical variables are not genuine degrees of freedom but mere spurious
variables devoid of physical meaning. The {\it observables} are not sensitive to the
values of these spurious variables, nor to the choice of the parametrization. The
super-Hamiltonian and super-momenta constraints of General Relativity are an example of
such a set of first class constraints.\cite{koxf,k2}

According to Dirac's method, the gauge invariance is preserved at the quantum level by
including in the Hilbert space only those states that are annihilated by the constraint
operators ({\it physical states}):

\begin{equation}
\hat {\cal H}_o \psi=0,\ \ \ \ \ \ \ \ \ \hat {\cal H}_a \psi=0, \label{1}
\end{equation}
and a linear operator must be inserted in the inner product to kill the integrations on
the spurious degrees of freedom (gauge fixing) .\cite{ht}

In order that the prescription (\ref{1}) be consistent with the algebra of first class
constraints, a proper operator ordering should be found out such that the constraint
algebra is realized at the quantum level in the following way:

\begin{equation}
[{\hat {\cal H}}_\alpha,{\hat {\cal H}}_\beta]={\hat C}^\gamma_{\alpha\beta}( q^{i},{\hat
p}_j){\hat {\cal H}}_\gamma \label{2}
\end{equation}
where $\alpha$, $\beta$ stand for both $o$ and $a$.

Such an operator ordering can be found out by raising the {\it BRST generator} $\Omega$ to
the status of a hermitian and nilpotent operator. The BRST generator is a fermionic
magnitude defined in a phase space extended by the addition of pair of canonically
conjugated variables $(\eta^\alpha,{\cal P}_\alpha)$ ({\it ghosts}) for each first class
constraint. Ghosts have parity opposed to the one of the respective constraint. $\Omega(q,
p, \eta, {\cal P})$ is defined by the conditions \cite{ht}

\begin{equation}
\{\Omega,\Omega\}=0,\ \ \ \ \ \ \ \ \ \ \Omega=\eta^\alpha{\cal H}_\alpha + more\label{22}
\end{equation}
(the Poisson bracket is symmetric for fermionic quantities). In Eq. (\ref{22}) ``more"
means terms of higher order in the ghosts. $\Omega$ is a conserved charge of the extended
system, and generates a global (rigid) symmetry. A hermitian and nilpotent realization of
$\Omega$ captures the structure of the algebra (\ref{2}) in the equation

\begin{equation}
0=[\hat \Omega,\hat \Omega]=2 \hat\Omega^2 \label{3}
\end{equation}

So, well ordered Dirac operator constraints can be identified from the form of
$\hat\Omega$. Also Dirac physical states can be mapped into a cohomological class of
physical states ($\hat\Omega\Psi=0$) of the nilpotent operator $\hat\Omega$. The BRST
extended system contains both classical and quantum behavior of the constrained system.

We will consider the following system:

\begin{equation}
{\cal H}_o={1\over 2} g^{ij}(q) p_i p_j+v, ~~~~~~{\cal H}_a=\xi_a^j(q) p_j, \label{vce}
\end{equation}
where $v$ is a gauge invariant potential (invariant under transformations generated by the
super-momenta), and $g^{ij}$ is an indefinite non-degenerate metric. The constraint
algebra is

\begin{equation}\{{\cal H}_o,{\cal H}_a\}=C_{oa}^{b}(q, p){\cal H}_b
=C_{oa}^{bj}(q)p_j {\cal H}_b, \label{cc}\end{equation}

\begin{equation}\{{\cal H}_a,{\cal H}_b\}=C_{ab}^{c}(q) {\cal
H}_c. \label{cc2}\end{equation} In this case the BRST generator is

\begin{equation}
\Omega=\eta^\alpha  {\cal H}_\alpha  + {1\over  2}  \eta^\alpha \eta^\beta
C_{\alpha\beta}^\gamma {\cal P}_\gamma ,\label{omegacl}
\end{equation}
A hermitian and nilpotent realization of $\Omega$ is \cite{prd}

\begin{equation}
\hat\Omega={\hat\Omega}^{linear}+{\hat\Omega}^{quad}\label{omeganil}
\end{equation}
where

\begin{equation}
{\hat\Omega}^{linear}=f^{1\over 2}\left[\eta^a\xi^i_a {\hat p}_i +{1\over 2}\eta^a\eta^b
C_{ab}^c \hat{\cal P}_c \right]f^{-{1\over 2}}\label{omeganil1}
\end{equation}
\begin{eqnarray}
{\hat\Omega}^{quad}=&\eta^{o}\left({1\over 2}f^{-{1\over 2}}{\hat p}_i f g^{ij}{\hat p}_j
f^{-{1\over 2}}+v\right)\cr\cr &+{1\over 2}f^{-{1\over 2}}{\hat p}_i f
\eta^{o}\eta^{a}C^{bi}_{oa}{\hat {\cal P}}_b f^{-{1\over 2}}\cr\cr &+{1\over 2}f^{-{1\over
2}}{\hat {\cal P}}_a f \eta^{o}\eta^{b}C^{aj}_{ob}{\hat p}_j f^{-{1\over
2}},\label{omeganil2}
\end{eqnarray}
and $f$ solves the equation

\begin{equation}
C^b_{ab}=f^{-1}(f\xi^i_a)_{,i}.\label{div}
\end{equation}
($f$ is a volume in the gauge orbit of the supermomenta).

In order to read from $\hat\Omega$ the constraint operators fulfilling Eq. (\ref{2}),
$\hat\Omega$ must be rearranged in $\eta-{\cal P}$ order by repeatedly using the ghost
(anti)-commutation relations. After this procedure is completed, the classical structure
of Eq. (\ref{omegacl}) will be reproduced at the quantum level \cite{ht}

\begin{equation}
{\hat\Omega}=\eta^\alpha  \hat{\cal H}_\alpha  +    {1\over  2} \eta^\alpha \eta^\beta
\hat{C}_{\alpha\beta}^\gamma {\hat{\cal P}}_\gamma.\label{omegacl2}
\end{equation}
In our case the result is

\begin{equation}
{\hat{\cal H}}_o=f^{1\over 2}\left[{1\over 2}  f^{-1} \hat p_i f g^{ij} \hat p_j + v +
{i\over 2} C_{oa}^{aj}\hat p_j \right] f^{-{1\over 2}}\label{hath}
\end{equation}
\begin{equation}
{\hat{\cal H}}_a= f^{1\over 2}\xi^i_a{\hat p}_i f^{-{1\over 2}}
\end{equation}
\begin{equation}
{\hat C}^b_{oa}={1\over 2}(f^{1\over 2} C_{oa}^{bj}\hat p_j f^{-{1\over 2}} + f^{-{1\over
2}} \hat p_j C_{oa}^{bj} f^{1\over 2} )
\end{equation}
\begin{equation}
{\hat C}^c_{ab}= C^c_{ab}\label{c}
\end{equation}

Although it was supposed that the potential $v$ is gauge invariant to render simpler the
constraint algebra, the results can be generalized to potentials that can be factorized as
a gauge invariant function $v$ times a never null function $\vartheta(q)$. This can be
accomplished by performing a unitary transformation leading to a different hermitian and
nilpotent BRST generator

\begin{equation}
\hat \Omega  \rightarrow  e^{i\hat  G}~\hat\Omega~ e^{-i\hat G}.\label{tu}
\end{equation}
By choosing
\begin{equation}\hat G={1\over 2}[\hat\eta^o~ ln~  \vartheta(q)~\hat{\cal
P}_o-\hat{\cal P}_o~ ln~ \vartheta(q)~\hat\eta^o],\label{de}
\end{equation}
the operators (\ref{hath}-\ref{c}) suffer the following changes:

\smallskip

$v\longrightarrow V = \vartheta\ v$,

\smallskip

$g^{ij}\longrightarrow G^{ij}= \vartheta\ g^{ij}$,

\smallskip

$f\longrightarrow \vartheta^{-1}\ f$,

\smallskip

$C^{bj}_{oa}\longrightarrow \vartheta\ C^{bj}_{oa}$,

\smallskip
\noindent and new structure functions appear:
\begin{equation}
C^o_{oa}=\xi^i_a (\ln\vartheta)_{,i}\ .
\end{equation}

Thus the unitary transformation (\ref{tu}-\ref{de}) amounts to the scaling of the
classical super-Hamiltonian constraint. The scaling of a constraint does not modify the
classical dynamics. The quantum system also remains unchanged provided that the Dirac
physical states change according to
\begin{equation}
\varphi  \rightarrow    \varphi'= \vartheta^{-1/2}\ \varphi.
\end{equation}

In order to define an inner product between Dirac physical states, the spurious variables
associated with the super-momenta constraints should be frozen by the ordinary procedure
of inserting the Dirac deltas of the gauge fixing functions and the corresponding
Fadeev-Popov determinant. However the reparametrization invariance associated with the
super-Hamiltonian constraint should be managed in a more specific way. If the potential is
positive definite, then the system can be deparametrized as a relativistic particle, where
the time is hidden in the configuration space ({\it intrinsic time}); time essentially is
a canonical variable inside the light-cone of the metric $g^{ij}$. Therefore a variable
behaving as {\it time} (i.e., monotonically increasing on every classical trajectory)
should be also frozen together with the rest of the spurious variables.\footnote[1]{This
is not the case for General Relativity, where the potential is the spatial curvature
$^3R$}

Things can be not so straightforward in other cases. As an example, let us consider the
case where the potential is not definite positive but there exists a time-like vector
$\vec\xi_o$ such that \cite{fs2}

\begin{equation}
{\cal L}_{\vec\xi _o}(|\vec\xi_o|^{-2}\ V)= 1\label{condgb'}
\end{equation}
and
\begin{equation}
{\cal L}_{\vec\xi _o}(|\vec\xi_o|^{-2}\ {\bf G})\approx 0\label{condggb'}
\end{equation}
(i.e. $\vec\xi_o$ is a Killing vector of the scaled metric $|\vec\xi_o|^{-2}\ {\bf G}$ on
the constraint surface). Then, one could factorize out the function
$\vartheta=|\vec\xi_o|^2$ from the super-Hamiltonian to get a simpler although equivalent
constraint:

\begin{equation}
{\cal L}_{\vec\xi _o} v = 1\label{condgb'2}
\end{equation}
and
\begin{equation}
{\cal L}_{\vec\xi _o}{\bf g}\approx 0\label{condggb'2}
\end{equation}
${\sqrt 2}\ \vec\xi_o$ is a unitary Killing vector for the scaled metric ${\bf g}$. In a
coordinate system where the parameter of $\vec\xi_o$ is the $q^o$ coordinate, and the rest
of the coordinate basis --$\{\partial /\partial q^\mu\}$-- is orthogonal to $\vec\xi_o$,
the former geometrical properties become

\begin{equation}
\vec\xi _o = {\partial\over\partial q^o}\ ,\label{conddgb'}
\end{equation}

\begin{equation}
{\partial v\over\partial q^o} = 1\label{condgb'3}
\end{equation}
and
\begin{equation}
{\partial g^{ij}\over\partial q^o}\approx 0\ .\label{condggb'3}
\end{equation}
Then the super-Hamiltonian looks
\begin{equation}
{\cal H}_o=-{1\over 2}p_o^2+{1\over 2} g^{\mu\nu}(q^\lambda) p_{\mu}p_{\nu}+{\cal
V}(q^{\mu})+q^o  \label{ham}
\end{equation}

where ${\cal V}(q^\mu)$ is a gauge invariant potential. One could deparametrize this
system by performing the following canonical transformation

\begin{equation}
q^o=p_t ,~~~~~~~~~~~~~~~  p_o=-t \label{tc}
\end{equation}
Thus the super-Hamiltonian becomes
\begin{equation}
{\cal H}_o=p_t+{1\over 2} g^{\mu\nu} p_{\mu}p_{\nu}+{\cal V}(q^{\mu})-{1\over 2}t^2
\label{ham2}
\end{equation}
This is nothing but the Hamiltonian constraint of a trivially parametrized system. Coming
back to the original variables, one realizes that the variable to be frozen in the inner
product is not a coordinate but a momenta ({\it extrinsic time} \cite{york}) \cite{bf}.
This ``gauge fixing" should be done in a non conventional way. However, the transformation
(\ref{tc}) in the above examined example teach us that the required insertion in the inner
product involves an integral operator performing a Fourier transform \cite{fs2}.

\bigskip

This work was supported by Universidad de Buenos Aires (Proy. TX 64) and Consejo Nacional
de Investigaciones Cient\'\i ficas y T\'ecnicas.

\end{document}